\documentclass[10pt, conference]{IEEEtran}
\IEEEoverridecommandlockouts

\usepackage{cite}
\usepackage{amsmath,amssymb,amsfonts}
\usepackage{algorithmic}
\usepackage{graphicx}
\usepackage{textcomp}
\usepackage{xcolor}
\usepackage{multirow}
\usepackage[export]{adjustbox}
\usepackage{lipsum}
\usepackage{hyperref}

\def\BibTeX{{\rm B\kern-.05em{\sc i\kern-.025em b}\kern-.08em
    T\kern-.1667em\lower.7ex\hbox{E}\kern-.125emX}}

\begin{document}

\title{Area Optimization of Open-Source Low-Power INA in 130nm CMOS using Hybrid Mixed-Variable PSO
\vspace{-30pt}
}

\author{
\IEEEauthorblockN{Avishka Herath\textsuperscript{1},
Chanula Luckshan\textsuperscript{1},
Lochana Katugaha\textsuperscript{2},
Udara Mendis\textsuperscript{3} and
Kithmin Wickremasinghe\textsuperscript{4}} \\
\IEEEauthorblockA{$^{1}$Department of Electronic and Telecommunication Engineering, University of Moratuwa, Sri Lanka}
\IEEEauthorblockA{$^{2}$Department of Electrical and Electronic Engineering, Sri Lanka Institute of Information Technology, Sri Lanka}
\IEEEauthorblockA{$^{3}$Hamburg University of Technology, Germany}
\IEEEauthorblockA{$^{4}$Department of Electrical and Computer Engineering, University of British Columbia, Canada}
\vspace{-20pt}
}

\maketitle

\begin{abstract}
As open-source silicon initiatives democratize access to integrated circuit development using multi-project environments, silicon area has become a premium resource. However, minimizing this layout area traditionally forces designers to compromise on core performance specifications. To address this challenge, this paper presents an open-source framework based on a hybrid mixed-variable particle swarm optimization algorithm and the $g_m/I_D$ methodology to minimize the layout area of complex analog circuits while meeting design requirements. The framework's efficacy is demonstrated by designing a low-power instrumentation amplifier that achieves a $90.33\%$ reduction in gate area over existing implementations.
\end{abstract}


\section{Introduction}

Analog circuit design is a complex process that has traditionally relied heavily on the designers' intuition and experience. This is primarily due to the highly nonlinear relationship between circuit performance and design parameters \cite{razavi2016design}. Recently, the emergence of open-source silicon initiatives, such as Tiny Tapeout \cite{venn2024tt}, has democratized access to application-specific integrated circuit development. However, in these shared, multi-project environments, silicon area is a premium resource that strictly limits the physical feasibility of a design. Consequently, rigorous area optimization has shifted from being a secondary consideration to a primary economic constraint for open-source analog CMOS design.

\begin{figure}[t!]
    \centering
    \includegraphics[width=0.95\linewidth]{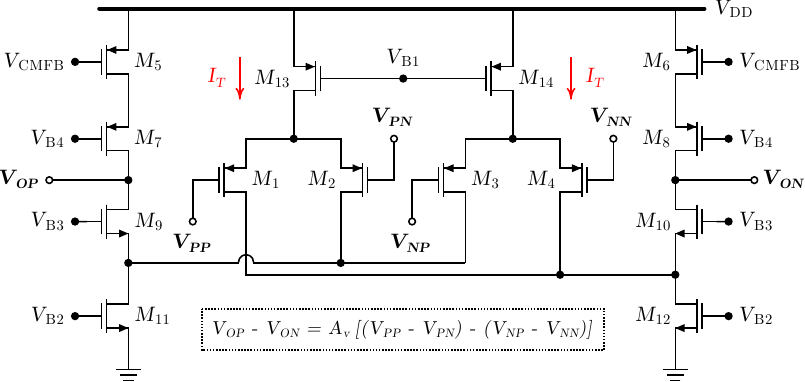}
    \vspace{-0.2cm}
    \caption{Schematic of an FDDA-based INA at the transistor level.}
    \vspace{-0.4cm}
    \label{fig:ina}
\end{figure}

To reduce dependence on manual sizing and accelerate the design cycle, automated design has been pursued by formulating circuit sizing as a nonlinear constrained optimization problem \cite{rashid2022pso}. While gradient-descent and convex optimization techniques exist, evolutionary algorithms (EAs) like particle swarm optimization (PSO) have proven highly effective \cite{rashid2022pso}. 
Although automated sizing via EAs is well-documented for fundamental building blocks like the 5-transistor operational transconductance amplifier, scaling these methodologies to larger, multi-stage systems remains a significant challenge. For example, practical biomedical applications demand highly sophisticated blocks like the fully differential difference amplifier (FDDA)-based instrumentation amplifier (INA) shown in Fig.~\ref{fig:ina}. However, this topology introduces a vast search space for transistor parameters, requiring simultaneous tuning of the core amplifier, common-mode feedback (CMFB) circuit, and the bias networks \cite{adornes2025fdda}. 

In this paper, we present an automated, area-optimized design of a complex FDDA-based INA utilizing a hybrid mixed-variable PSO (HMV-PSO) algorithm. 
The results confirm PSO's scalability to moderately large analog systems, reducing manual design effort while establishing a robust methodology for deploying high-performance, area-efficient analog macros in open-source tapeout platforms.

\section{Optimization Problem Overview}
\label{sec: optimization problem}

Analog CMOS design is inherently a multidimensional optimization problem where improving one parameter often comes at the direct expense of another \cite{razavi2016design}.
\begin{figure*}[t!]
    \centering
    \includegraphics[width=1\linewidth]{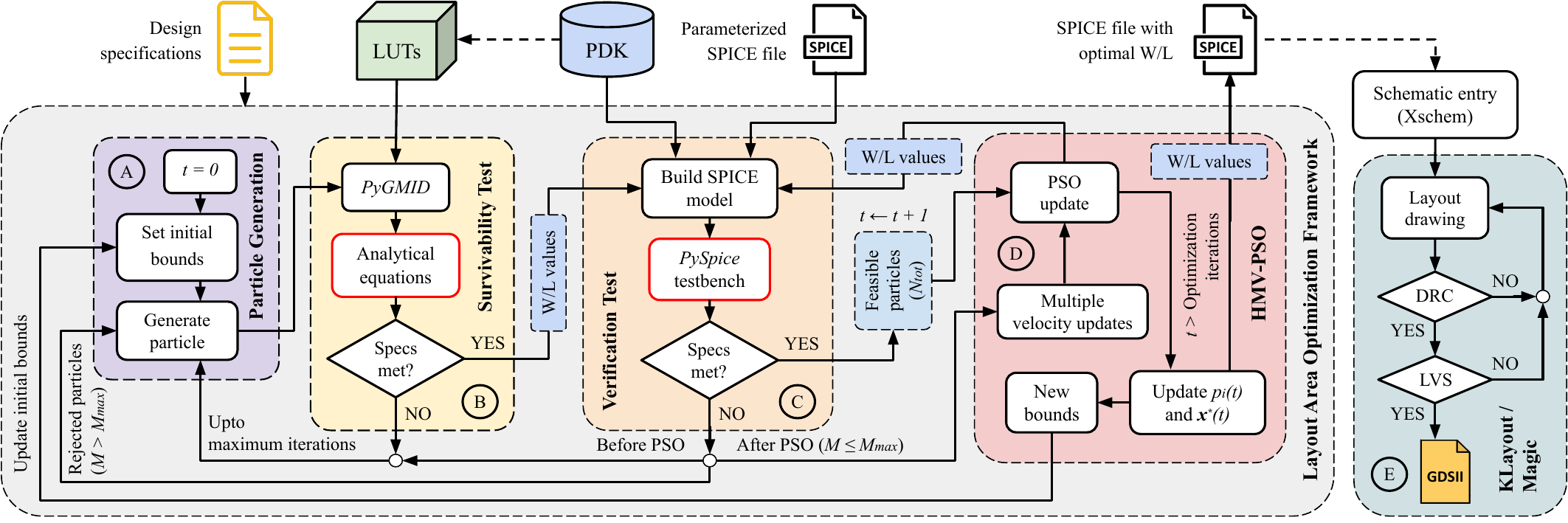}
    \vspace{-0.6cm}
    \caption{\textbf{Overview of the proposed HMV-PSO optimization framework for layout area minimization.} Given design specifications, LUT data, PDK models, and a parameterizable SPICE netlist as inputs, the flow proceeds through: (A) particle generation within adaptively bounded search spaces; (B) analytical feasibility filtering via PyGMID; (C) full-circuit SPICE verification via PySpice; and (D) iterative PSO-based position updates cycling through (A)–(C) until convergence on the area-optimal sizing. (E) The resulting solution is used for the standard physical design flow to produce the GDSII layout.}
    \vspace{-0.4cm}
    \label{fig: overall framework}
\end{figure*}

\subsection{\texorpdfstring{$g_m/I_D$}{gm/ID} Primer using SKY130A Process Design Kit (PDK)}

To systematically navigate this complex trade-off space, the $g_m/I_D$ methodology is widely adopted, enabling faster optimization of the initial design \cite{jespers_murmann_2017, 10798363}.
Unlike traditional square-law models, this approach relies on transconductance efficiency ($g_m/I_D$) as the primary design parameter \cite{SABRY201887} and maps device behavior via pre-characterized lookup tables (LUTs) generated from the target PDK.
Our work is implemented in the open-source SKY130A process, utilizing the publicly available $g_m/I_D$ starter kit \cite{murmann_gmid_sky130}.

\subsection{Problem Formulation}

For a given CMOS circuit, the independent design variables typically consist of $g_m/I_D$, the channel length ($L$), and the bias current ($I_{D}$) for each transistor. Once these three variables are selected, the required device width ($W$) can be deterministically extracted from the LUTs \cite{jespers_murmann_2017}. 
However, the folded-cascode architecture of the FDDA in Fig.~\ref{fig:ina} requires strict symmetry to ensure proper differential operation. And to enforce this, we assume that $M_{1,2,3,4}$, $M_{5,6}$, $M_{7,8}$, $M_{9,10}$, $M_{11,12}$, and $M_{13,14}$ in Fig.~\ref{fig:ina} are matched.
We denote the bias tail current of the input pairs as $I_T$ and set the cascode branch starving current to $I_T/4$, allowing the amplifier to achieve a higher DC gain and power efficiency \cite{razavi2016design}.
Under these constraints, the majority of the transistor currents become dependent variables, reducing the effective design space to $N=6$ independently sizable transistor groups.

\subsection{Design Variables, Objective Function, and Constraints}

The primary objective is to minimize the total chip area while meeting all design specifications.
The gate area, computed over all transistor groups, is therefore adopted as the fitness function.
The position vector, $\boldsymbol{x}$, and the fitness function, $f(\boldsymbol{x})$, are defined as follows:
\begin{equation}
\label{eq: position vector}
\boldsymbol{x} = [g_m / I_{D,1}, L_1, \ldots , g_m / I_{D,N}, L_N, I_T],
\end{equation}
\begin{equation}
f(\boldsymbol{x}) = \sum_{i=1}^{N} (W_i \cdot L_i),
\end{equation}
where $W_i$ and $L_i$ are the width and length of the $i$-th transistor group, respectively.
Also, it should be noted that LUTs are typically generated for a pre-defined set of $L$ values. Due to second-order effects \cite{razavi2016design}, transistor characteristics do not scale linearly with $L$. Consequently, $L$ is treated as a discrete variable that must be selected directly from this pre-defined set, while the other variables are treated as continuous.

To ensure the circuit meets functional requirements, specifications such as DC voltage gain ($A_{v\text{0}}$), unity-gain bandwidth (GBW), phase margin, slew rate, common-mode rejection ratio (CMRR), power supply rejection ratio (PSRR), and power dissipation are defined as optimization constraints. 

\section{Optimization Methodology}
\label{sec: framework}

In this work, we propose an HMV-PSO framework to solve the problem formulated in Sec.~\ref{sec: optimization problem}. The overall flow is illustrated in Fig.~\ref{fig: overall framework}.

\subsection{Particle Generation}

The swarm is initialized by generating a set of particles, each representing a candidate design vector $\boldsymbol{x} \in \mathbb{R}^{13}$, as defined in \eqref{eq: position vector}. Continuous variables are sampled within analytically determined maximum bounds for $g_m/I_{D,i}$ and $I_T$ based on performance constraints, while discrete variables are drawn from a predefined set of $L$ values.
As defining a large search space for $g_m/I_D$ increases computational runtime, we employ an adaptive bounding strategy:

\subsubsection{Initial Bound Selection}

The designer specifies a narrow initial range for each $g_m/I_D$ variable based on the required channel inversion level of its corresponding transistor group.

\subsubsection{Dynamic Bound Adjustment}

As the optimization progresses, the search bounds for $g_m/I_D$ variables are dynamically tightened whenever a new best position is discovered, accelerating the convergence. Specifically, upon finding a new  position 
$\boldsymbol{x}^* $, the lower 
and upper bounds for each $g_m/I_D$ variable are updated as:
\begin{equation}
    \left[ g_m/I_{D,i}^* - \Delta,\; g_m/I_{D,i}^* + \Delta \right], 
    \quad i = 1, \dots, 6,
\end{equation}
where $\Delta$ is a fixed shrinkage margin. 
This adjustment progressively focuses the search on the neighborhood of the most promising region, minimizing wasted evaluations in unpromising regions.

\subsection{Survivability Test (PyGMID)}

After generating a particle, the algorithm tests whether it lies within the optimization problem's feasible region.
For each vector $\boldsymbol{x}$, the lookup functions of the open-source Python $g_m/I_D$ toolkit PyGMID \cite{pygmid} are used to extract the transistor parameters necessary for analytically evaluating the performance metrics and verifying whether the candidate particle satisfies the design goals in Table~\ref{tab:FDDA}.
If a particle fails this test, it is discarded and replaced with a newly generated design vector. 
This process repeats until the swarm is fully populated with feasible particles.
As this test uses only analytically derived equations and LUT data, it serves as an efficient preliminary filter that rapidly eliminates undesired particles at a minimal computational cost.

\subsection{Verification Test (PySpice)}

When a particle passes the initial survivability test or updates its position during an iteration, it undergoes a full circuit simulation to verify whether the design goals in Table~\ref{tab:FDDA} are met. These simulations are executed in Ngspice (version 44), using the Python wrapper library PySpice \cite{PySpice}. The $W$ and $L$ values for each transistor, along with the bias voltages, are calculated using transistor parameters extracted via LUTs and passed as circuit variables to dynamically generate a SPICE netlist using a parameterizable template. Following this, the necessary analyses are performed across relevant testbenches to extract and evaluate the performance metrics. Particles that satisfy all constraints are retained; otherwise, they undergo the recovery procedure detailed in Sec.~\ref{sec: recovery}

\subsection{Particle Swarm Optimization for Mixed-Variable Problems}

At each iteration, particle positions are updated using distinct reproduction mechanisms for continuous and discrete variables, following the $\text{PSO}_{mv}$ formulation of \cite{WANG2021100808}.

\subsubsection{Continuous Reproduction Method}

For continuous variables, the standard PSO velocity and position update rules 
are employed. At each iteration, the velocity of the $i$-th particle is updated 
as:
\begin{align}
    v_i(t+1) &= w \cdot v_i(t) + c_1(t) \cdot r_1 \cdot \bigl(p_i(t) - x_i(t)\bigr)\nonumber \\ 
    &\hspace{30pt}+ c_2(t) \cdot r_2 \cdot \bigl(\boldsymbol{x}^*(t) - x_i(t)\bigr),
    \label{eq:velocity}
\end{align}
\begin{equation}
\label{eq: position update}
x_i(t+1) \leftarrow x_i(t) + v_i(t+1),
\end{equation}
where $w$ is the inertia weight, $c_1(t)$ and $c_2(t)$ are the time-varying cognitive and social acceleration coefficients respectively, $r_1, r_2 \sim  \mathcal{U}(0,1)$ are independent random scalars drawn at each update, $p_i(t)$ is the personal best position of particle $i$, and $\boldsymbol{x}^*(t)$ is 
the global best position at iteration $t$. Velocities are initialized to  $\pm 10\%$ of the respective variable's bound range to avoid large initial displacements.

\subsubsection{Discrete Reproduction Method}

For each discrete variable $j$, a probability distribution $Prob_{j,n}(t)$ tracks the likelihood of assigning the $n$-th available value, initialized uniformly as $Prob_{j,n}(0) = \frac{1}{\hat{n}_j}$, where $\hat{n}_j$ is the number of available values for variable $j$.

During each iteration, among the total population of $N_{tot}$ particles, the half with the lowest personal best areas is used to update the probability distributions:
\begin{equation}
Prob_{j,n}(t+1) = \alpha \times Prob_{j,n}(t) + (1-\alpha) \times \frac{Count_{j,n}}{N_{tot}/2},
\end{equation}
where $Count_{j,n}$ is the number of particles in the superior half that carry the $n$-th value for variable $j$, and $\alpha$ is a parameter that balances historical and current search information. New discrete values are sampled independently according to the updated distribution. This mechanism biases future samples toward the length values that have proven most effective among the swarm's elite, while $\alpha$ prevents premature collapse of the distribution.

\subsection{Adaptive Parameter Selection}

To balance exploration and exploitation throughout the optimization, the $c_1$ and $c_2$ coefficients in \eqref{eq:velocity} are varied linearly over the course of the run:
\begin{equation}
    c_k(t) = c_{k,\max} - \frac{(c_{k,\max} - c_{k,\min})}{T_{\max}} \cdot t \quad k=1,2 \quad,
\end{equation}
where where $T_{\max}$ is the total number of iterations. As $t$ increases, $c_1$
decreases and $c_2$ increases, gradually shifting each particle's motion from self-guided exploration toward collective convergence around the global best. 

\subsection{Infeasible Particle Recovery}
\label{sec: recovery}
Because the continuous velocity update in \eqref{eq:velocity} can move particles into regions that fail either the survivability test or the full SPICE verification test, a structured recovery procedure is applied to any particle that is rejected at the verification stage.

Up to $M_{\max}$ additional velocity updates are performed for the rejected particle using \eqref{eq:velocity} and \eqref{eq: position update}, with each candidate offspring evaluated through both the survivability and verification tests. If a feasible offspring is found within these $M_{\max}$ attempts, it replaces the rejected particle, and the recovery is declared successful.

\section{Results and Discussions}

The framework described in Sec.~\ref{sec: framework} was configured with the following parameters: $w=0.5$, $\Delta =1$, $c_{k,\max}=2$, $c_{k,\min}=1$, $M_{\max}=5$, and $\alpha=0.7$. A swarm size of 20 particles was employed. These values were selected based on a preliminary study that evaluated convergence speed and solution quality across the representative parameter sweep. A comprehensive ablation study is deferred to an extended version of this work due to space constraints.

\subsection{Optimization Results}

\begin{figure}[!t]
    \centering
    \includegraphics[width=1\columnwidth]{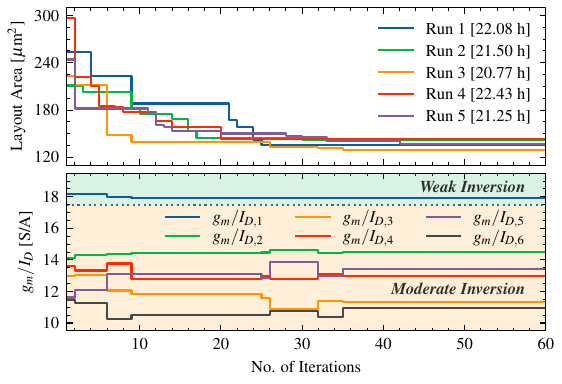}
    \vspace{-0.6cm}
    \caption{HMV-PSO convergence over 60 iterations: (top) layout area reduction across 5 runs; (bottom) $g_m/I_{D,i}$ trajectories and inversion levels for Run 3.}
    \vspace{-0.4cm}
    \label{fig: pso results}
\end{figure}

Fig.~\ref{fig: pso results} (top) illustrates the convergence profiles and execution times of five independent PSO runs, each conducted over 60 iterations on a machine with an AMD Ryzen 7 5800H processor and 16 GB of RAM. The best fitness value across all runs is retained as the final solution, and the corresponding circuit design parameters are given in Table~\ref{table: sizing parameters}. The average run time for the design was recorded to be 21.61 hours.
\begin{table}[b!]
\renewcommand{\arraystretch}{1.2}
\vspace{-0.4cm}
\caption{Optimal sizing parameters for the FDDA circuit (Run 3)}
\vspace{-0.1cm}
\label{table: sizing parameters}
\centering
\begin{tabular}{c c c c}
\hline
\hline
\textbf{Transistor} & \textbf{W / L}\ ($\mu$m) & \textbf{Transistor} & \textbf{W / L}\ ($\mu$m) \\
\hline
M\textsubscript{1} - M\textsubscript{4} & 75.64 / 0.3 & M\textsubscript{9} - M\textsubscript{10} & 0.66 / 3.0 \\
M\textsubscript{5} - M\textsubscript{6} & 0.84 / 0.4 & M\textsubscript{11} - M\textsubscript{12} & 2.48 / 2.0 \\
M\textsubscript{7} - M\textsubscript{8} & 0.69 / 1.0 & M\textsubscript{13} - M\textsubscript{14} & 2.55 / 0.7  \\
\hline
\end{tabular}
\end{table}

The optimizer effectively converges the $g_m/I_{D,i}$ values shown in Fig.~\ref{fig: pso results} (bottom), automatically settling the input pairs in the weak inversion region to maximize intrinsic gain~\cite{jespers_murmann_2017}, while driving the remaining transistors into the moderate inversion region for optimal low-power operation~\cite {dorrer2025open}.

\subsection{Layout Design and Post-Layout Simulation Results}

\begin{figure}[!h]
    \centering
    \vspace{-0.1cm}
    \includegraphics[width=1\linewidth]{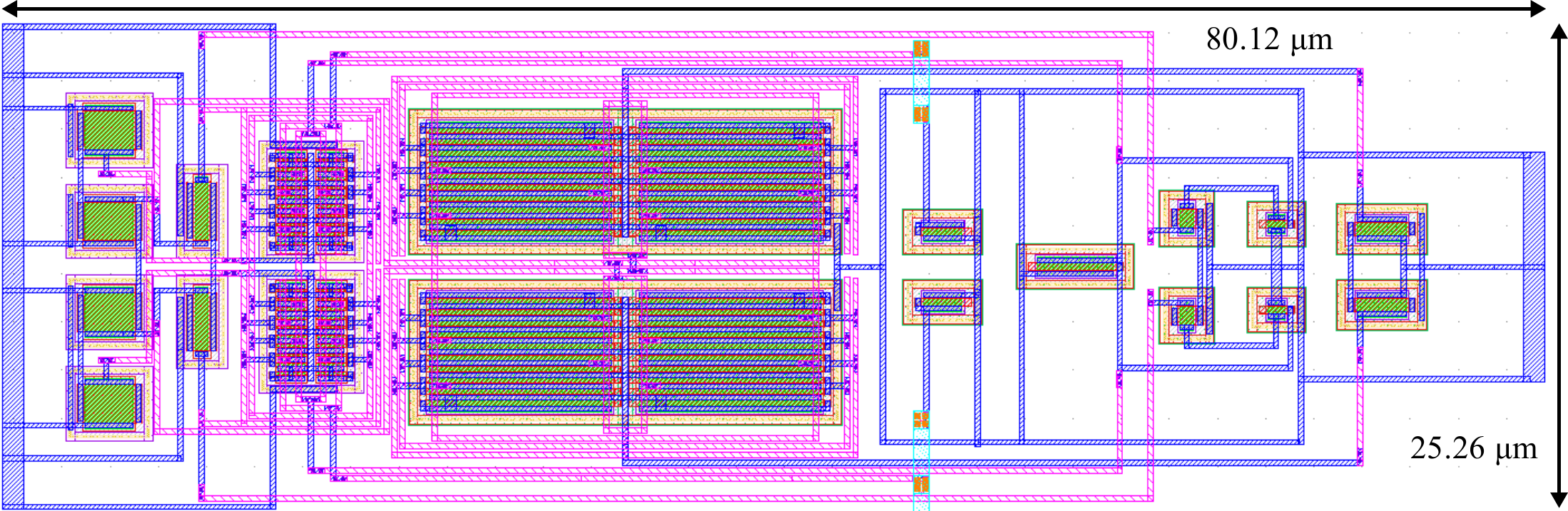}
    \vspace{-0.5cm}
    \caption{Layout of the FDDA and CMFB circuit ($80.12\ \mu$m $\times$ $25.26\ \mu$m).}
    \label{fig:layout}
\end{figure}

To ensure accurate characterization, CMFB is integrated with FDDA during optimization. Bias voltages were applied using ideal voltage sources during simulations. The resulting layout, designed with Table~\ref{table: sizing parameters} sizings, is shown in Fig.~\ref{fig:layout}.

\begin{table}[!b]
\centering
\renewcommand{\arraystretch}{1.2} 
\vspace{-0.4cm}
\caption{Performance comparison of the designed FDDA circuit.}
\vspace{-0.1cm}
\label{tab:FDDA}
\begin{tabular}{lclcc}
\hline
\hline
\multirow{2}{*}{\textbf{Parameter}} & \textbf{Adornes,} & \multicolumn{1}{c}{\textbf{Design}} & \multicolumn{2}{c}{\textbf{Our Work}} \\ \cline{4-5} 
 & \textbf{et al. \cite{adornes2025fdda}} & \multicolumn{1}{c}{\textbf{Goals}} & \textbf{Pre-layout} & \textbf{Post-layout} \\
\hline
$A_{v\text{0}}$ [dB] & 72 & $\geq$ 72 & 72.33 & 72.21 \\ 
GBW [MHz] & 47.77 & $\geq$ 1 & 1.05 & 0.95 \\ 
Phase margin [$^\circ$] & 55.46 & $\geq$ 60 & 84.81 & 81.92 \\ 
Slew rate [V/$\mu$s] & 6.54 & $\geq$ 1 & 1.02 & 0.99 \\ 
CMRR [dB] @ 1 kHz & 119.9 & $\geq$ 120 & 203.83 & 80.10 \\ 
PSRR [dB] @ 1 kHz & 67.49 & $\geq$ 60 & 226.18 & 84.05 \\ 
Power\textsuperscript{*} [$\mu$W] & 219.6 & $\leq$ 40 & 19.65 & 19.62 \\ 
$C_L$\textsuperscript{+} [pF] & 0.25 & \multicolumn{1}{c}{1} & 1 & 1 \\
Gate area [$\mu$m$^2$] & 1140 & \multicolumn{1}{c}{\textit{min.}} & 110.27 & 110.32 \\ \hline
\end{tabular}
\begin{flushleft}
\textsuperscript{*} Power consumption calculated as $V_{\text{DD}} \times I_{\text{DD}}$ where $V_{\text{DD}} = 1.8V$. \\
\textsuperscript{+} Differential capacitive load; $2 \times C_L$ is connected to a single-ended output. 
\end{flushleft}
\end{table}

\begin{figure}[t!]
    \centering
    \includegraphics[width=1\linewidth]{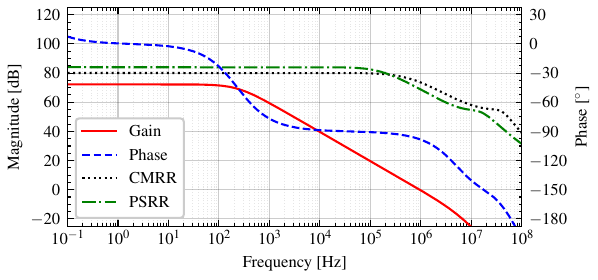}
    \vspace{-0.6cm}
    \caption{Post-layout simulation results of the FDDA circuit: frequency response plot showing the gain, phase, CMRR, and PSRR after parasitic extraction.}
    \vspace{-0.4cm}
    \label{fig:freq_response}
\end{figure}

This FDDA design requires a total gate area of only 110.27 $\mu$m$^2$. This yields a substantial $90.33\%$ area reduction when evaluated against the similar architecture presented in \cite{adornes2025fdda}. Pre-layout simulation results obtained directly from the optimizer, alongside the post-layout results, are summarized in Table~\ref{tab:FDDA}, demonstrating that core design specifications are met. However, the GBW and slew rate were strategically relaxed to better suit low-frequency biomedical applications. The systems' frequency response can be observed in Fig.~\ref{fig:freq_response}.

\section{Conclusions and Future Works}

This paper presented an automated, open-source framework that leverages an HMV-PSO algorithm combined with the $g_m/I_D$ methodology to aggressively minimize the layout area of complex analog CMOS designs. Demonstrated on an FDDA-based INA in the SKY130A process, the optimizer effectively converges to the optimal operating region for each transistor as required by the design specifications. Because \textit{PySpice} currently does not support noise spectrum simulations, future work will focus on integrating noise as a formal design specification. Additionally, subsequent optimizations will account for the effect of transistor fingers to ensure proper matching during layout, thereby improving differential operation and post-layout CMRR.

\bibliographystyle{IEEEbib}
\bibliography{refs}

\end{document}